\documentclass[aps,pre,twocolumn,a4paper]{revtex4}

\usepackage{graphicx}
\usepackage{amsmath}
\usepackage[dvips]{color}

\begin{document}

\title{Accurate determination of crystal structures based on averaged local bond order parameters}

\author{Wolfgang Lechner and Christoph Dellago}

\affiliation{Faculty of Physics, University of Vienna, Boltzmanngasse 5, 1090 Vienna, Austria}

\date{\today}

\begin{abstract}
Local bond order parameters based on spherical harmonics, also known as Steinhardt order parameters, are often used to determine crystal structures in molecular simulations. Here we propose a modification of this method in which the complex bond order vectors are averaged over the first neighbor shell of a given particle and the particle itself. As demonstrated using soft particle systems, this averaging procedure considerably improves the accuracy with which different crystal structures can be distinguished.
\end{abstract}

\maketitle

\section{Introduction}

In computational studies of crystallization from the undercooled liquid one needs to be able to distinguish particles that are part of the crystal from those that belong to the liquid. Ideally, such an assignment should be based on the local environment of the particles only. One method to do that, which is independent of the specific crystal structure and does not require the definition of a reference frame, is provided by the following algorithm based on spherical harmonics \cite{FRENKEL_NUC}. First, the complex vector $q_{lm}(i)$ of particle $i$ is defined as \cite{STEINHARDT}
\begin{equation}
\label{equ:localbondordervector}
q_{lm}(i) =  \frac{1}{N_b(i)} \sum_{j=1}^{N_b(i)} Y_{lm} ({\bf r}_{ij}).
\end{equation}
Here, $N_b(i)$ is the number of nearest neighbors of particle $i$, $l$ is a free integer parameter and $m$ is an integer that runs from $m=-l$ to $m=+l$. The functions $Y_{lm}({\bf r}_{ij})$ are the spherical harmonics and ${\bf r}_{ij}$ is the vector from particle $i$ to particle $j$. Using the set of complex vectors $q_{6m}$ one can then define the scalar product 
\begin{equation}
\label{equ:scalarproduct}
S_{ij} = \sum_{m=-6}^{6}q_{6m}(i)q_{6m}^*(j),
\end{equation}
which measures the correlation between the structures surrounding particles $i$ and $j$. The $*$ indicates complex conjugation. Two particles $i$ and $j$ are defined to be \textit{connected} if $S_{ij}$ is larger than a given value, typically $S_{ij}>0.5$. A particle is solid-like if the number of connections it has with its neighbors is above a certain threshold, typically between $6$ and $8$. If a particle is connected to less particles, it is considered to be liquid-like. Using this criterion to distinguish solid-like from liquid-like particles one can then search for clusters of connected solid-like particles. The size of the largest of these crystalline clusters $\hat{N}$ is often used as a reaction coordinate to follow the progress of the crystallization process. Provided that this reaction coordinate captures the essential physics of the crystal nucleation, the Gibbs free energy $G(\hat{N})$ calculated as a function of $\hat{N}$ provides a means to estimate the rate at which the nucleation of the crystalline phase occurs.

This procedure very efficiently distinguishes between solid-like and liquid-like particles, but does not discriminate between different crystal structures. A set of parameters which hold the information of the local structure are the local bond order parameters, or Steinhardt order parameters, defined as \cite{STEINHARDT}
\begin{equation}
\label{equ:localbondorderparameter}
q_l(i) = \sqrt{ \frac{4 \pi}{2l+1} \sum_{m=-l}^{l} |q_{lm}(i)|^2}.
\end{equation}
Depending on the choice of $l$, these parameters are sensitive to different crystal symmetries. Each of them depends on the angles between the vectors to the neighboring particles only and therefore these parameters are independent of a reference frame. Different approaches based on these local bond order parameters were developed to analyze the structure of the crystalline nucleus during the freezing event. Especially $q_4$ and $q_6$ are often used as they are a good choice to distinguish between cubic and hexagonal structures \cite{MORONI_BOLHUIS,Coasne,OGATA}. 

In practice, local bond order parameters are used in different ways. Frenkel and coworkers \cite{WOLDE_MONTERO,WOLDE_FRENKEL,MORONI_BOLHUIS,Volkov} analyzed the structure of crystalline clusters in terms of order parameter distributions. To do that, they first computed the distributions of $q_4$ and $q_6$ for the liquid and for perfect FCC and BCC crystals. Due to thermal fluctuations, these distributions can be rather broad. Then, they determined the distributions of the same order parameters in the crystalline cluster. These distributions are represented as a superposition of the distribution functions of the perfect phases. The superposition coefficients $c_{\rm FCC}$, $c_{\rm BCC}$ and $c_{\rm LIQ}$, determinded by mean square minimization, then yield information on the composition of the cluster. For instance, $c_{\rm FCC} \approx 0.5$, $c_{\rm BCC} \approx 0.5$,and $c_{\rm LIQ} \approx 0$ would be indicative of a cluster that is half in the FCC structure and half in the BCC structure.

Another bond order parameter method, used for instance in Ref. \cite{DESGRANGES}, defines regions in the two dimensional $q_4$-$q_6$-plane and $q_4$-$w_4$-plane. The crystalline structure around a given particle is characterized by its positions in these planes. The parameter $w_l$ necessary for that analysis is defined as 
\begin{equation}
\label{equ:wl}
w_l(i) =  \frac{\sum\limits_{m_1+m_2+m_3=0} \begin{pmatrix} l & l & l \\ m_1 & m_2 & m_3 \end{pmatrix}q_{lm_1}(i)q_{lm_2}(i)q_{lm_3}(i)}{\left(\sum\limits_{m=-l}^{l}|q_{lm}(i)|^2\right)^{3/2}}.
\end{equation}
Here, the integers $m_1$, $m_2$ and $m_3$ run from $-l$ to $+l$, but only combinations with $m_1+m_2+m_3=0$ are allowed. The term in brackets is the Wigner 3-$j$ symbol \cite{LANDAU_LIFSCHITZ_QM}. Using this approach, one can determine the type of crystalline structure occurring around each individual particle.

As mentioned above, thermal fluctuations smear out the order parameter distributions such that it may be difficult to distinguish local crystalline structures based on Steinhardt bond order parameters. In the following we present a simple method to increase the accuracy of the crystal structure determination by averaging over the bond order parameters of nearest neighbor particles. This averaging procedure is discussed in Sec. \ref{sec:averaging} and validated for two different soft sphere systems in Sec. \ref{sec:results}. Some conclusions are provided in Sec. \ref{sec:conclusions}.

\section{Averaged Bond  Order Parameters}
\label{sec:averaging}

The crystal structure determination described above can be improved by using the following averaged form of the local bond order parameters:
\begin{equation}
\label{equ:averagedlocalbondorderparameter}
\bar{q}_l(i) = \sqrt{ \frac{4 \pi}{2l+1} \sum_{m=-l}^{l} |\bar{q}_{lm}(i)|^2},
\end{equation}
where 
%
%
%
\begin{equation}
\label{equ:averagedlocalbondordervector}
\bar{q}_{lm}(i) =  \frac{1}{\tilde{N}_b(i)} \sum_{k=0}^{\tilde{N}_b(i)}q_{lm}(k).
\end{equation}
Here, the sum from $k = 0$ to $\tilde{N}_b(i)$ runs over all neighbors of particle $i$ plus the particle $i$ itself. Thus, to calculate $\bar{q}_l(i)$ of  particle $i$ one uses the local orientational order vectors $q_{lm}(i)$ averaged over particle $i$ and its surroundings. While $q_l(i)$ holds the information of the structure of the first shell around particle $i$, its averaged version $\bar{q}_l(i)$ also takes into account the second shell. One might say that using the parameter $\bar{q}_l$ instead of $q_l$ increases the accuracy of the distinction of different structures at the price of a coarsening of the spacial resolution. In this sense, the averaged local bond order parameter is similar to the scalar product of Eq. (\ref{equ:scalarproduct}) used to decide whether a particle is in a solid-like or liquid-like environment. Also in that case the second particle shell is effectively taken into account.\\ 
The averaged orientational order parameter $\bar{q}_{lm}$ can also be used to define an averaged version $\bar{w}_l$ of the order parameter $w_l$,
\begin{equation}
\label{equ:awl}
\bar{w}_l(i) =  \frac{\sum\limits_{m_1+m_2+m_3=0} \begin{pmatrix} l & l & l \\ m_1 & m_2 & m_3 \end{pmatrix}\bar{q}_{lm_1}(i)\bar{q}_{lm_2}(i)\bar{q}_{lm_3}(i)}{\left(\sum\limits_{m=-l}^{l}|\bar{q}_{lm}(i)|^2\right)^{3/2}}.
\end{equation}

\section{Results}
\label{sec:results}

In the following, we verify that these averaged forms of the bond order parameters indeed increase the accuracy of the crystal structure determination. The calculations are done for two distinct systems: one in which the particles interact via a Lennard-Jones potential and one in which the interaction is of the Gaussian core form \cite{STILLINGER_GCM,GAUSSIAN_CORE_PHASE,GAUSSIAN_SOFTLY_PHASE}. 

\subsection{Lennard-Jones system}

For the calculations of the Lennard-Jones system \cite{LJPHASE} the temperature was $k_{\rm B}T=0.92 \epsilon$ and the pressure was $P=5.68 \epsilon\sigma^{-3}$, corresponding to 20\% undercooling of the liquid phase \cite{WOLDE_MONTERO}. The same conditions were used in Ref. \cite{WOLDE_MONTERO} to study homogeneous crystal nucleation. This phase point corresponds to a mean density of $\rho \simeq 1.05 \sigma^{-3}$ in the FCC crystal, which is the most stable phase under these conditions. All simulations were carried out in the isobaric-isothermal ensemble, but calculations at constant volume for the corresponding densities yielded essentially identical results. The particle number was $N=432$ in the BCC crystal, $N=864$ in the FCC crystal, and $N=512$ in the HCP crystal and in the undercooled liquid. Two particles were defined to be neighbors if their distance was smaller than $r_N = 1.4 \sigma$. 

\begin{figure}[htb,floatfix]
\centerline{\includegraphics[width=5.5cm]{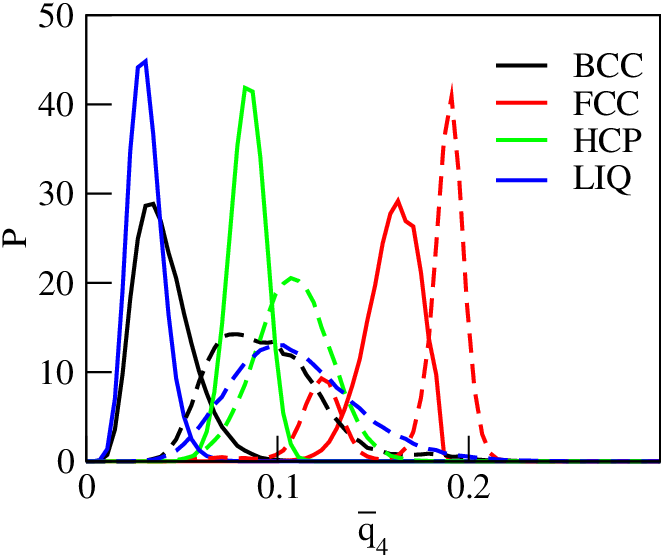}}
\centerline{\includegraphics[width=5.5cm]{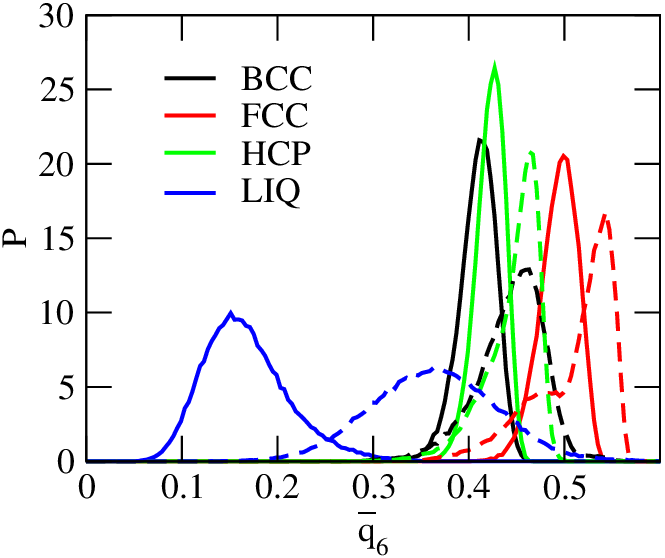}}
\caption{Top: Probability distributions of $\bar{q}_4$ (solid lines) and $q_4$ (dashed lines) for the FCC, BCC and HCP crystals and for the undercooled liquid (LIQ) in the Lennard-Jones system. Bottom: Probability distributions of $\bar{q}_6$ (solid lines) and $q_6$ (dashed lines) for the same phases. }
\label{fig:qhistolj}
\end{figure}

In Fig. \ref{fig:qhistolj} we compare the distributions of the local bond order parameters $q_4$ and $q_6$ with those of the averaged bond order parameters $\bar{q}_4$ and $\bar{q}_6$.  Due to the averaging procedure, the overlap between the distributions decreases considerably thus allowing one to distinguish between different structures more precisely. As can be inferred from Fig. \ref{fig:qhistolj}, the reduced overlap is due not only to a narrowing of the distributions but also to a shift of the means towards smaller values. While one would expect the order parameter distributions to narrow on the basis of the central limit theorem, one is tempted to attribute the shift of the means to local correlations. To test this speculation, we have artificially eliminated correlations by computing $\bar{q}_l$ from uncorrelated values of $q_{lm}$ generated with the appropriate statistics. Also in this case the narrowing of the distributions was accompanied by a shift of the means indicating that this shift is not due to local correlations between the local order parameters of nearest neighbors but rather originates in the functional form of $\bar{q}_l$.

\begin{table}[htb,floatfix]
\begin{tabular}{|c|cc|cc|}
\hline
&  $q_4$ &  $\bar{q}_4$  & $q_6$ &  $\bar{q}_6$  \\
\hline
BCC & 0.089988 & 0.033406 & 0.440526 & 0.408018 \\
FCC & 0.170880 & 0.158180 & 0.507298 & 0.491385 \\
HCP & 0.107923 & 0.084052 & 0.445384 & 0.421813 \\
LIQ & 0.109049 & 0.031246 & 0.360012 & 0.161962 \\
\hline
\end{tabular}
\caption{Mean of the distributions of $q_4$, $\bar{q}_4$, $q_6$, and $\bar{q}_6$ for the BCC, FCC, and HCP crystals and the undercooled liquid in the Lennard-Jones system. }
\label{tbl:meanlj}
\end{table}
\begin{table}[htb,floatfix]
\begin{tabular}{|c|cc|cc|}
\hline
 &  $q_4$ &  $\bar{q}_4$  & $q_6$ &  $\bar{q}_6$  \\
\hline
BCC & 0.026831 & 0.010782 & 0.034791 & 0.020516\\
FCC & 0.032787 & 0.014346 & 0.043301 & 0.020566\\
HCP & 0.019476 & 0.009434 & 0.028992 & 0.015965\\
LIQ & 0.031992 & 0.008786 & 0.066518 & 0.039360\\
\hline
\end{tabular}
\caption{Standard deviation of the distributions of $q_4$, $\bar{q}_4$, $q_6$, and $\bar{q}_6$ for the BCC, FCC, and HCP crystals and the undercooled liquid in the Lennard-Jones system. }
\label{tbl:sigmalj}
\end{table}

Table \ref{tbl:meanlj} and Table \ref{tbl:sigmalj} show the mean value and the standard deviation of the distributions of $q_4$, $\bar{q}_4$, $q_6$ and $\bar{q}_6$, respectively, for all four phases we studied. The standard deviation decreases by a factor of 2-4 in most cases. To quantify the improvement of the averaged order parameter we define the overlap between two distributions as
\begin{equation}
\label{equ:overlap}
O_{\alpha \beta}= \frac{1}{N_{\alpha \beta}} \int P_{\alpha}(x)P_{\beta}(x) \mathrm dx.
\end{equation}
Here, the indices $\alpha$ and $\beta$ denote the various crystal structures and $N_{\alpha \beta}$ is given by 
\begin{equation}
\label{equ:overlapnormalization}
N_{\alpha \beta}= \sqrt{\int P^2_{\alpha}(x) \mathrm dx \int P^2_{\beta}(y) \mathrm dy}.
\end{equation}
%
\begin{table}[htb,floatfix]
\begin{tabular}{|c|cc|cc|}
\hline
&  $q_4$ &  $\bar{q}_4$  & $q_6$ &  $\bar{q}_6$  \\
\hline
BCC-FCC   & 0.124377 & 0.000000 & 0.33959496 & 0.017147  \\
BCC-HCP   & 0.738776 & 0.007880 & 0.94732246 & 0.866026 \\
BCC-LIQ   & 0.881215 & 0.992978 & 0.46374455 & 0.000004 \\
FCC-HCP   & 0.210924 & 0.000188 & 0.31311693 & 0.033288 \\
FCC-LIQ   & 0.233250 & 0.000000 & 0.14914436 & 0.000000 \\
HCP-LIQ   & 0.938029 & 0.001096 & 0.35744697 & 0.000000 \\
\hline
\end{tabular}
\caption{Overlap between the distributions of $q_4$, $\bar{q}_4$, $q_6$, and $\bar{q}_6$ for the various phases in the Lennard-Jones system. }
\label{tbl:overlapseverallj}
\end{table}

\begin{figure}[htb,floatfix]
  \centering
  \begin{minipage}[b]{4.25 cm}
    \includegraphics[width=3.9cm]{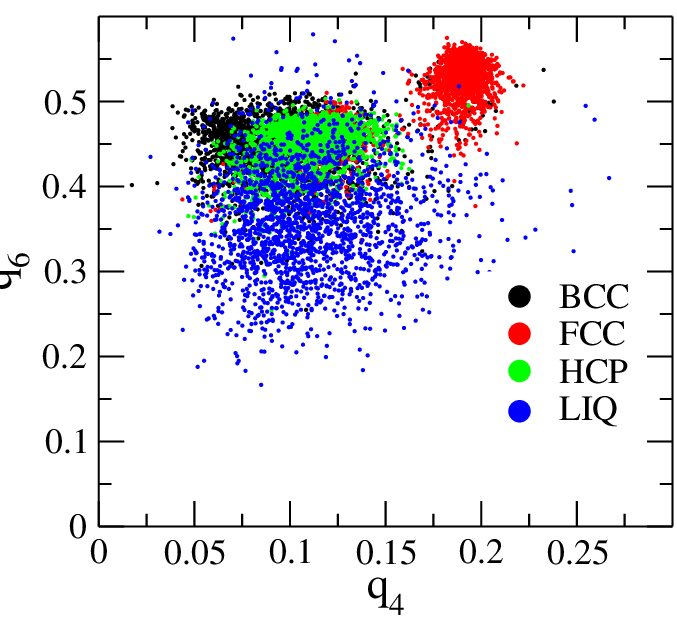}  
  \end{minipage}
  \begin{minipage}[b]{4.25 cm}
    \includegraphics[width=3.9cm]{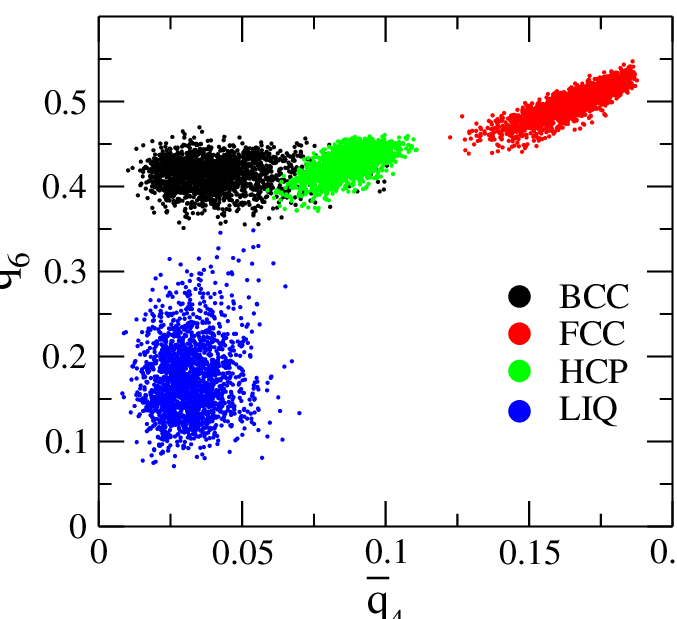} 
  \end{minipage}

  \hspace{0.1cm}\\

  \begin{minipage}[b]{4.25 cm}
    \includegraphics[width=3.9cm]{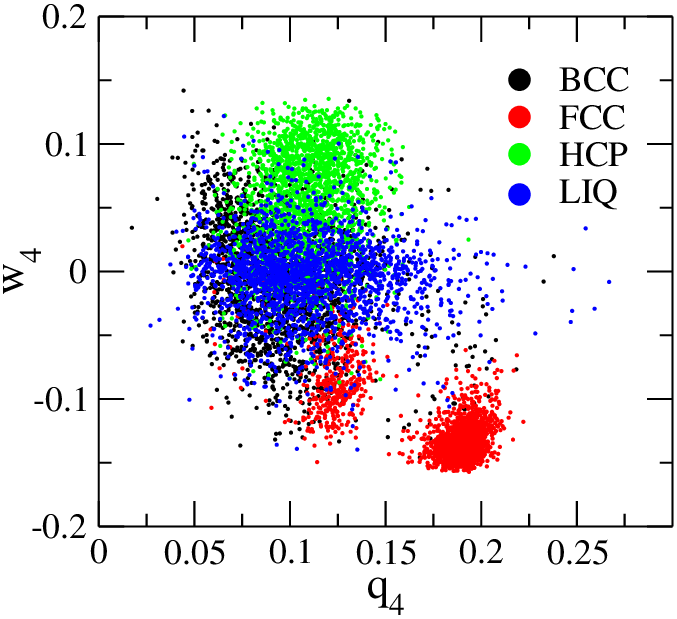}  
  \end{minipage}
  \begin{minipage}[b]{4.25 cm}
    \includegraphics[width=3.9cm]{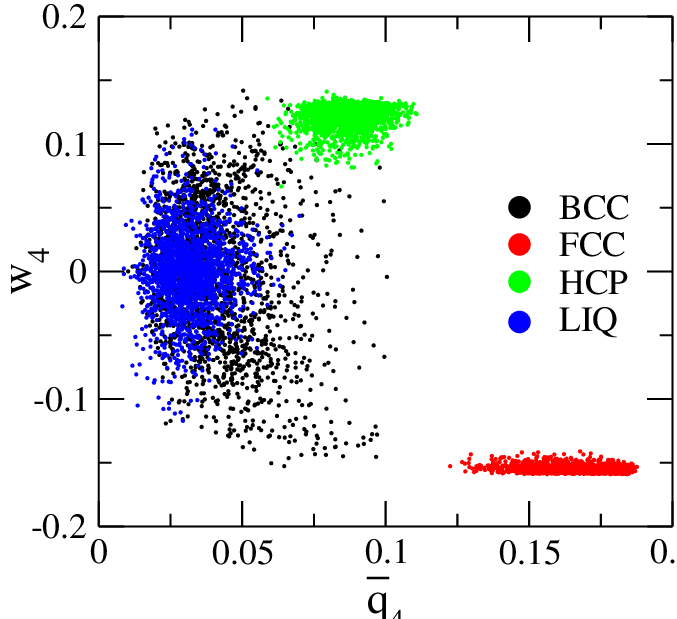} 
  \end{minipage}
  \caption{Top: Comparison between the $q_4$-$q_6$-plane (left) and the $\bar{q}_4$-$\bar{q}_6$-plane (right) for the Lennard-Jones system in three different crystalline structures and in the liquid phase. Each point corresponds to a particular particle, where $2000$ points from each structure were chosen randomly. Bottom: $q_4$-$w_4$-plane (left) and the $\bar{q}_4$-$\bar{w}_4$-plane (right).}
  \label{fig:ljq4q6}
\end{figure}

In Table \ref{tbl:overlapseverallj} we summarize the overlap between the distributions of $q_4$, $\bar{q}_4$, $q_6$, and $\bar{q}_6$ for BCC, FCC, and HCP crystals and the undercooled liquid. In most cases the overlap decreases by several orders of magnitude when going from the local order parameter to the averaged version. One exception are the $q_4$ distributions of the liquid and the BCC crystal, which have an overlap of nearly $1$ even for the averaged order parameter. All phases are, however, resolved very well using two averaged bond order parameters. Figure \ref{fig:ljq4q6} shows a comparison of the various crystalline and liquid phases as scatter plots in the $q_4$-$q_6$-plane, the $\bar{q}_4$-$\bar{q}_6$-plane, the $q_4$-$w_4$-plane, and the $\bar{w}_4$-$\bar{w}_4$-plane. For the averaged version of the order parameters (right) the four phases separate much better. In practice, the separation between BCC and HCP is the most difficult one. In the  $\bar{q}_4$-$\bar{q}_6$-plane these two crystal structures can be distinguished easily and also in the $\bar{q}_4$-$\bar{w}_4$-plane they are well separated. When going from $q_4$ to $\bar{q}_4$ the overlap between these two structures decreases by two orders of magnitude (see Tab. \ref{tbl:overlapseverallj}). 

\subsection{Gaussian Core Model}

For the Gaussian core system, in which particles interact via the pair-interaction $v(r) = \varepsilon \exp[-(r/\sigma)^2]$ \cite{STILLINGER_GCM,GAUSSIAN_CORE_PHASE,GAUSSIAN_SOFTLY_PHASE}, we chose a temperature of $k_B T = 0.0033\varepsilon$ and pressure $P=0.011\varepsilon\sigma{-3}$ at which the thermodynamically most stable phase is an FCC crystal with density $0.11\sigma^{-3}$. These conditions correspond to the same level of undercooling as that of the Lennard-Jones system described above. All calculations were carried out in the isothermal-isobaric ensemble and the particle numbers for the various crystal structures where the same as in the Lennard-Jones system. The nearest neighbor distance was $r_N = 3.0\sigma$.

\begin{figure}[htb,floatfix]
\centerline{\includegraphics[width=5.5cm]{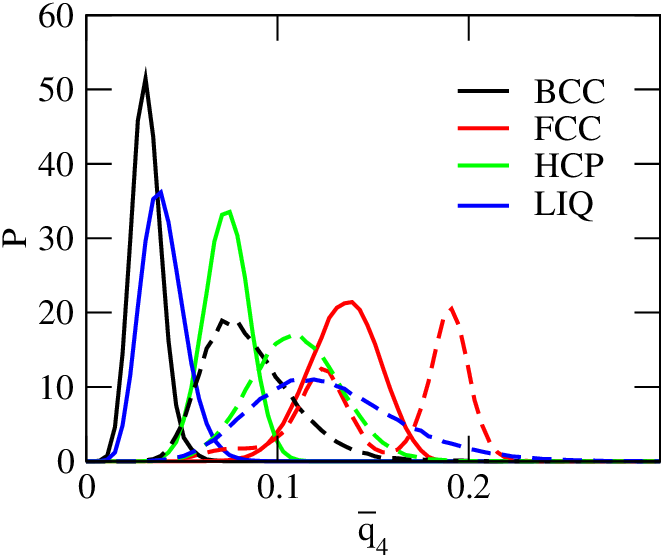}}
\centerline{\includegraphics[width=5.5cm]{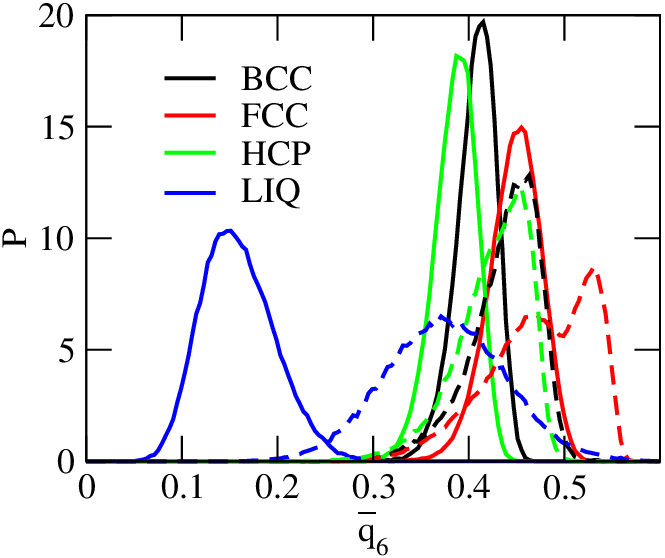}}
\caption{Top: Probability distributions of $\bar{q}_4$ (solid lines) and $q_4$ (dashed lines) for the FCC, BCC and HCP crystals and for the undercooled liquid (LIQ) in the Gaussian core system. Bottom: Probability distributions of $\bar{q}_6$ (solid lines) and $q_6$ (dashed lines) for the same phases.  }
\label{fig:qhisto}
\end{figure}

\begin{figure}[htb,floatfix]
  \centering
  \begin{minipage}[b]{4.25 cm}
    \includegraphics[width=3.9cm]{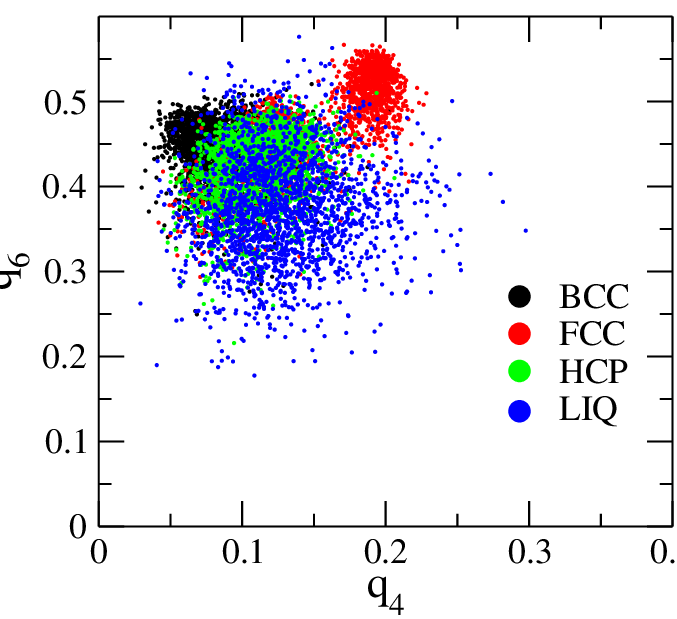}  
  \end{minipage}
  \begin{minipage}[b]{4.25 cm}
    \includegraphics[width=3.9cm]{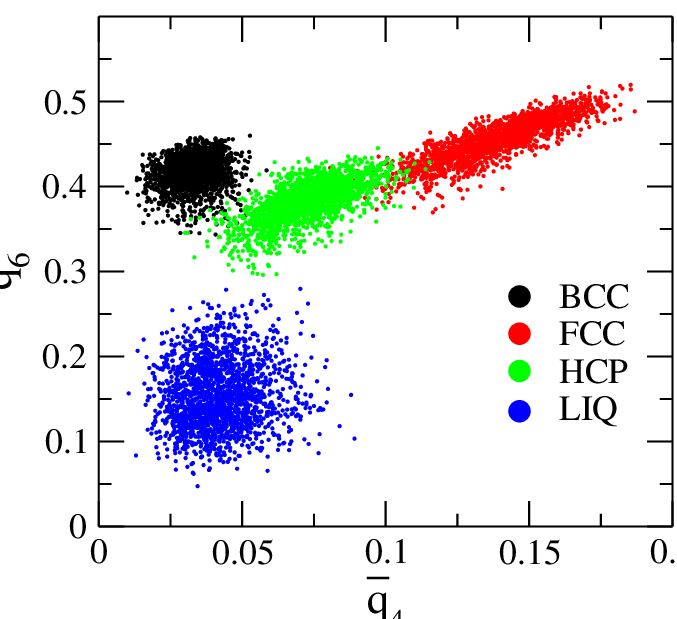} 
  \end{minipage}

  \hspace{0.1cm}\\

  \begin{minipage}[b]{4.25 cm}
    \includegraphics[width=3.9cm]{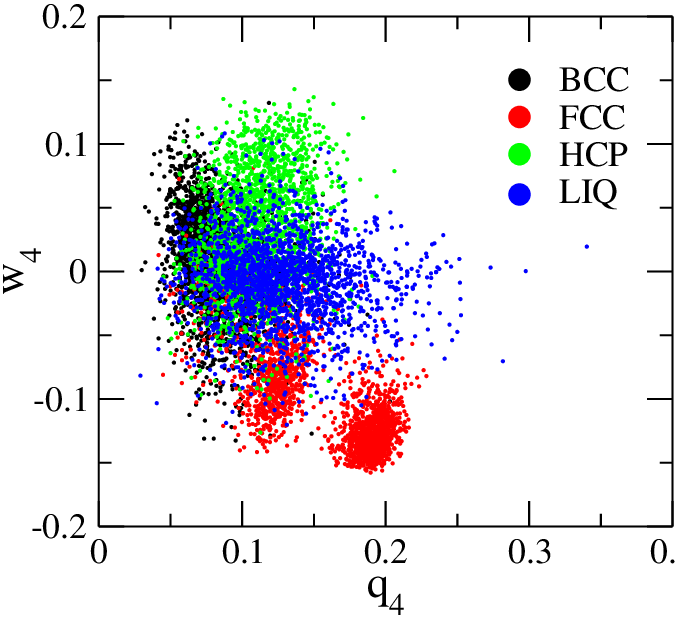}  
  \end{minipage}
  \begin{minipage}[b]{4.25 cm}
    \includegraphics[width=3.9cm]{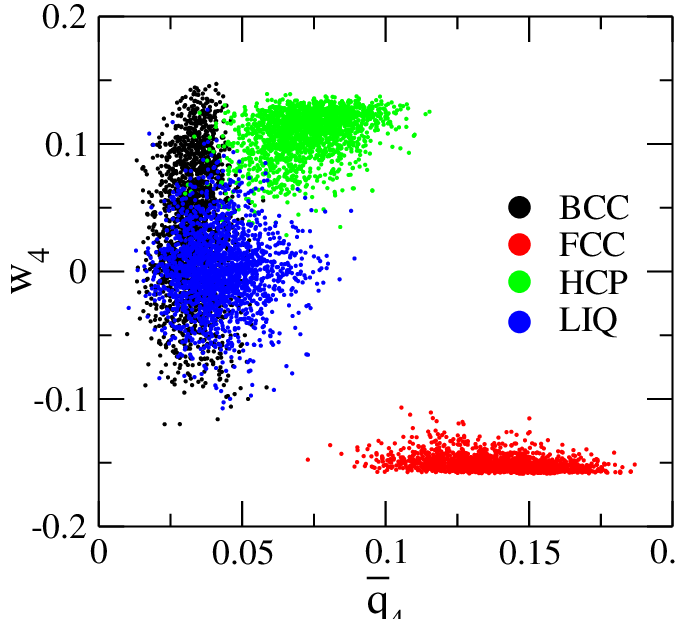} 
  \end{minipage}
  \caption{Top: Comparison between the $q_4$-$q_6$-plane (left) and the $\bar{q}_4$-$\bar{q}_6$-plane (right) in the Gaussian core system . Bottom: $q_4$-$w_4$-plane (left) and the $\bar{q}_4$-$\bar{w}_4$-plane (right).}
  \label{fig:gcq4q6}
\end{figure}

Also for the Gaussian core system we find a shift in the mean and a decrease in the width of the distributions (see Fig. \ref{fig:qhisto}). The mean and the standard deviation of the distributions for the various phases are listed in Tables \ref{tbl:mean} and \ref{tbl:sigma}, respectively. The overlap between the different structures is shown in Table \ref{tbl:overlapseveral}. In most cases the overlap decreases by a large factor, only the overlap in $q_4$ between BCC and liquid increases. But as in the Lennard-Jones case  all structures can be distinguished very well if two order parameters are taken into account (see Fig. \ref{fig:gcq4q6}).

\begin{table}[htb,floatfix]
\begin{tabular}{|c|cc|cc|}
\hline
 &  $q_4$ &  $\bar{q}_4$  & $q_6$ &  $\bar{q}_6$  \\
\hline
BCC & 0.085581 & 0.031728 & 0.437129 & 0.407515\\
FCC & 0.155336 & 0.134388 & 0.474079 & 0.447782\\
HCP & 0.109723 & 0.073369 & 0.424627 & 0.385720\\
LIQ & 0.126950 & 0.040297 & 0.375121 & 0.158913\\

\hline
\end{tabular}
\caption{Mean of the distributions of $q_4$, $\bar{q}_4$, $q_6$, and $\bar{q}_6$ for the BCC, FCC, and HCP crystals and the undercooled liquid in the Gaussian core system. }
\label{tbl:mean}
\end{table}

\begin{table}[htb,floatfix]
\begin{tabular}{|c|cc|cc|}
\hline
 &  $q_4$ &  $\bar{q}_4$  & $q_6$ &  $\bar{q}_6$  \\
\hline
BCC & 0.023841 & 0.008024 & 0.037274 & 0.021006\\
FCC & 0.039924 & 0.018200 & 0.053852 & 0.027126\\
HCP & 0.023617 & 0.011605 & 0.038462 & 0.022251\\
LIQ & 0.037891 & 0.011282 & 0.062874 & 0.038372\\
\hline
\end{tabular}
\caption{Standard deviation $\sigma$ of the distributions of $q_4$, $\bar{q}_4$, $q_6$, and $\bar{q}_6$ for the BCC, FCC, and HCP crystals and the undercooled liquid in the Gaussian core system. }
\label{tbl:sigma}
\end{table}

\begin{table}[htb,floatfix]
\begin{tabular}{|c|cc|cc|}
\hline
&  $q_4$ &  $\bar{q}_4$  & $q_6$ &  $\bar{q}_6$  \\
\hline
BCC-FCC   & 0.248341 & 0.000007 & 0.685253 & 0.462343 \\
BCC-HCP   & 0.687131 & 0.015436 & 0.964792 & 0.753398 \\
BCC-LIQ   & 0.619532 & 0.832443 & 0.581065 & 0.000006 \\
FCC-HCP   & 0.496870 & 0.022218 & 0.624299 & 0.197885 \\
FCC-LIQ   & 0.632466 & 0.000261 & 0.445508 & 0.000000 \\
HCP-LIQ   & 0.928871 & 0.128086 & 0.682437 & 0.000018 \\
\hline
\end{tabular}
\caption{Overlap between the distributions of $q_4$, $\bar{q}_4$, $q_6$, and $\bar{q}_6$ for the various phases in the Gaussian core system.}
\label{tbl:overlapseveral}
\end{table}

\section{Conclusions}
\label{sec:conclusions}

The averaged bond order parameters proposed in this paper separate different crystal structures more accurately than regular bond order parameters. Instead of using the first shell of surrounding particles only, this method also takes into account the second shell, just as one usually does when distinguishing between solid-like and liquid-like particles (see Eq. (\ref{equ:scalarproduct})). Due to the averaging procedure, the overlap between the order parameter distributions belonging to different phases decreases by up to a few orders of magnitude. The sharper distinction between the different phases is obtained at the cost of a slightly reduced spatial resolution. We demonstrated the enhancement in the crystal structure determination for two systems of soft particles interacting via a Lennard-Jones potential and a Gaussian core potential. For the same degree of undercooling the improvement is similar in both cases. Particularly in the $\bar{q}_4$-$\bar{q}_6$-plane the separation of the liquid phase from the solid phases is pronounced indicating that it might be possible to use this method also to distinguish between liquid and solid particles if only BCC, FCC and HCP structures are expected to be important for the process under study. We expect that the averaging procedure described in this paper lead to similar improvements also for other Steinhardt bond order parameters, e.g., $q_8$ and $w_8$. 

\begin{acknowledgments}
This research was supported by the University of Vienna through the University Focus Research Area {\em Materials Science} (project ``Multi-scale Simulations of Materials Properties and Processes in Materials'').
\end{acknowledgments}
 


\begin{thebibliography}{99}


\bibitem{FRENKEL_NUC}
S. Auer and D. Frenkel,
{\em Adv. Poly. Sci.},  {\bf 173}, 149 (2005).

\bibitem{STEINHARDT} 
P. Steinhardt, D. R. Nelson, and M. Ronchetti,
{\em Phys. Rev. B} {\bf 28}, 784 (1983).


\bibitem{MORONI_BOLHUIS} 
D. Moroni, P. R. ten Wolde, and P. G. Bolhuis, 
{\em Phys. Rev. Lett.} {\bf 94}, 235703 (2005).

\bibitem{Coasne} 
B. Coasne, S. K. Jain, L. Naamar, and K. E. Gubbins,
{\em Phys. Rev. B} {\bf 76}, 085416 (2007).

\bibitem{OGATA} 
Ogata, Shuji,
{\em Phys. Rev. A} {\bf 45}, 1122 (1992).

\bibitem{WOLDE_MONTERO} 
P. R. ten Wolde, M. J. Ruiz-Montero and D. Frenkel, 
{\em J. Chem. Phys.} {\bf 104}, 9932 (1996).

\bibitem{WOLDE_FRENKEL} 
P. R. ten Wolde and D. Frenkel, 
{\em Phys. Chem. Chem. Phys.} {\bf 1}, 2191 (1999).

\bibitem{Volkov} 
I. Volkov, M. Cieplak, J. Koplik, and J. R. Banavar,
{\em Phys. Rev. E} {\bf 66}, 061401 (2002). 

\bibitem{DESGRANGES} 
C. Desgranges and J. Delhommelle, 
{\em Phys. Rev. B} {\bf 77}, 054201 (2008).

\bibitem{LANDAU_LIFSCHITZ_QM} 
L. Landau and E. Lifschitz, 
``Quantum Mechanics'', Pergamon, London (1965).

\bibitem{STILLINGER_GCM} 
F. H. Stillinger, 
{\em J. Chem. Phys.} {\bf 65}, 3968 (1976).

\bibitem{GAUSSIAN_CORE_PHASE} 
S. Prestipino, F. Saijta and P. V. Giaquinta, 
{\em Phys. Rev. E} {\bf 71}, 050102(R) (2005).

\bibitem{GAUSSIAN_SOFTLY_PHASE} 
S. Prestipino, F. Saijta and P. V. Giaquinta, 
{\em J. Chem. Phys.} {\bf 123}, 144110 (2005).

\bibitem{LJPHASE} 
J. L. Johnson, J. A. Zollweg, and K. E. Gubbins,
{\em Mol. Phys.} {\bf 78}, 591 (1993). 


\end{thebibliography}
\end{document}